\documentclass[twocolumn,amsmath,amssymb,pre]{revtex4}

\usepackage{graphicx}
\usepackage{bm}
\usepackage{amssymb}
\usepackage{amsmath}
\usepackage{amsfonts}

\usepackage[utf8]{inputenc}

\begin{document}

\title{Modulational instability and collapse of internal gravity waves in the atmosphere}

\author{Volodymyr M. Lashkin}
\email{vlashkin62@gmail.com} \affiliation{$^1$Institute for
Nuclear Research, Pr. Nauki 47, Kyiv 03028, Ukraine}
\affiliation{$^2$Space Research Institute, Pr. Glushkova 40 k.4/1,
Kyiv 03187,  Ukraine}

\author{Oleg K. Cheremnykh}
\affiliation{Space Research Institute, Pr. Glushkova 40 k.4/1,
Kyiv 03187, Ukraine}

%\date{\today}

\begin{abstract}
Nonlinear two-dimensional internal gravity waves (IGWs) in the
atmospheres of the Earth and the Sun are studied. The resulting
two-dimensional nonlinear equation has the form of a generalized
nonlinear Schr\"{o}dinger equation with nonlocal nonlinearity,
that is when the nonlinear response depends on the wave intensity
at some spatial domain. The modulation instability of IGWs is
predicted, and specific cases for the Earth's atmosphere are
considered. In a number of particular cases, the instability
thresholds and instability growth rates are analytically found.
Despite the nonlocal nonlinearity, we demonstrate the possibility
of critical collapse of IGWs due to the scale homogeneity of the
nonlinear term in spatial variables.
\end{abstract}

\maketitle

\section{Introduction}

The spectrum of acoustic-gravity waves in the atmosphere of
planets and the Sun consists of acoustic and internal gravity
waves \cite{Hines1960,Tolstoy1967}, as well as evanescent wave
modes \cite{Waltercheid2005,Cheremnykh2019,Cheremnykh2021}.
Internal gravity waves, which are the lower branch of
acoustic-gravity atmospheric waves, have been intensively studied
in the physics of the Earth's and the Sun's atmosphere for more
than 60 years
\cite{Hines1960,Tolstoy1967,Liu1974,Beer1974,Gossard1975}. Space
missions have provided an additional incentive to study these
waves also in the atmospheres of other planets, for example, Mars
and Venus \cite{Schubert1984,Forbes2009}. Interest in IGWs is
largely due to the important contribution that these waves make to
the dynamics and energetics of the atmospheres of planets and the
Sun, ensuring effective redistribution of disturbance energy on a
global scale. In the Earth's atmosphere, these waves can be
generated by various sources of natural and anthropogenic origin.
In particular, IGWs are associated with the sources localized in
upper atmosphere and on the Sun, for example, precipitation of
charged particles at high latitudes, ionospheric currents, solar
terminator, etc. \cite{Prikryl2005,Fritts2008,Bespalova2016}. In
addition, IGWs in the upper atmosphere and ionosphere caused by
tropospheric or ground-based sources are currently being
intensively studied \cite{Rapoport2004,Yigit2008}. The linear
theory of IGWs has been developed in great detail (see, e.g.,
\cite{Sutherland2015}, reviews \cite{Francis1975,Fritts2003} and
references therein). In particular, the Coriolis force caused by
rotation, the presence of the magnetic field of the Earth and the
Sun \cite{Kaladze2008}, the effect of random temperature
inhomogeneity \cite{Lashkin2023}, etc. were taken into account. As
IGWs propagate upward in the atmosphere, their amplitudes rapidly
increase with altitude due to an exponential decrease in the
background density \cite{Hines1960,Liu1974,Roy2019}. In this
connection, when considering such waves, it becomes necessary to
take into account nonlinear effects. Nonlinear effects during the
propagation of IGWs have been studied in a number of works (see
also Ref.~\cite{Miropolsky2001} and references therein). In
particular, based on fluid equations with a term taking into
account force of gravity and an adiabatic equation of state,
nonlinear equations for IGWs in the atmosphere were derived in
Refs.~\cite{Stenflo1987,Stenflo1990,Stenflo2009}. The three-wave
interaction of IGWs and nonlinear responses were considered in
\cite{Dong1988,Fritts1992,Huang1991}. Interaction of atmospheric
gravity solitary waves with ion acoustic solitary waves was
studied in Ref.~\cite{Huang1992}. Nonlinear structures in the form
of convective cells of IGW waves in the Earth's atmosphere
\cite{Fedun2013}, tripole vortices and vortex chains
\cite{Jovanovic2001,Jovanovic2002}, dust-acoustic gravity vortices
\cite{Shukla1998}, and the so-called dust devils (rotating columns
of rising dust) \cite{Fedun2016,Fedun2021} were considered.
Nonlinear IGWs waves in a weakly ionized atmosphere in the form of
dipole vortices (cyclone-anticyclone pairs) were found in
Refs.~\cite{Misra2022IEEE,Misra2022AdvSpace}. In a recent paper
\cite{Lashkin2024}, nonlinear equations were obtained to describe
the dynamics of IGWs using the reductive perturbation method. In
the one-dimensional case, the corresponding solutions are
presented in the form of breather solitons, rogue waves, and dark
solitons. Some aspects of nonlinear IGW, including intensive
numerical modeling, have also been studied in
Refs.~\cite{Huang2014,Fritts2015,Snively2017,Fritts2019}.

In this paper, we study two-dimensional (2D) nonlinear IGWs based
on equations obtained with the aid of the reductive perturbation
method proposed in Ref.~\cite{Lashkin2024}. Unlike other works on
nonlinear atmospheric IGWs, we are the first to use this method to
derive a new 2D nonlinear equation with a nonlocal nonlinearity
when the nonlinear response depends on the wave packet intensity
at some extensive spatial domain. This equation may be treated as
a generalized nonlinear Schr\"{o}dinger equation, but the linear
part is essentially anisotropic, and moreover, the corresponding
linear operator can be either elliptic or hyperbolic. This
equation, in contrast to the original fluid equations describing a
stratified atmosphere, is quite adequately amenable to analytical
analysis and, despite its model nature, predicts the modulation
instability of IGWs and the possibility of wave collapse.

The 2D nonlinear Schr\"{o}dinger (NLS) equation with nonlocal
nonlinearity was considered in a number of works. An important
property of spatially nonlocal nonlinear response is that it
prevents a catastrophic collapse of multidimensional wave packets
which usually occurs in local self-focusing media with a cubic
nonlinearity. In particular, a rigorous proof of absence of
collapse in the model of the nonlocal NLS equation with
sufficiently general symmetric real-valued response kernel was
presented in Refs.~\cite{Turitsyn1985,Krolikovski2004}. It was
shown that nonlocal nonlinearity arrests the collapse and results
in the existence of stable coherent structures, not only the
fundamental soliton (which can collapse in the NLS with local
nonlinearity), but also dipole solitons, the so-called azimuthal
solitons (azimuthons) and vortex solitons
\cite{Lashkin2006,Torner2006,Lashkin2007PLA,Lashkin2007POP,Lashkin2007PRA}.
It is important that in these models the term with nonlocal
nonlinearity is not scale homogeneous in spatial variables. This
is precisely the reason for the absence of collapse. Despite the
nonlocal nonlinearity in our model, we predict the possibility of
IGW collapse, similar to the collapse of Langmuir waves in a
plasma \cite{Zakharov1972,Zakharov_UFN2012} and self-focusing of
nonlinear beams in optics \cite{Kivshar_book2003}. Collapse (since
the model is two-dimensional, collapse is critical) is possible
due to the fact that the nonlocal nonlinearity under consideration
is scale homogeneous in spatial variables. We also study the
modulation instability of IGWs, which is a precursor of collapse
that occurs at the nonlinear stage of instability.

The paper is organized as follows. In Sec. \ref{Sec2} the model
two-dimensional nonlinear equations for IGWs are presented. The
modulation instability is studied in Sec. \ref{Sec3}. In Sec.
\ref{Sec4} we demonstrate the possibility of collapse of IGWs. The
conclusion is made in Sec. \ref{Sec5}.

\section{\label {Sec2} Model equations}

Two-dimensional (2D) Stenflo equations
\cite{Stenflo1987,Stenflo2009} to govern the dynamics of nonlinear
atmospheric IGWs have the form
\begin{gather}
\frac{\partial}{\partial
t}\left(\Delta\psi-\frac{1}{4H^{2}}\psi\right)+\{\psi,\Delta\psi\}+\frac{\partial
\chi}{\partial x}=0, \label{main1}
\\
\frac{\partial \chi}{\partial
t}+\{\psi,\chi\}-\omega_{g}^{2}\frac{\partial \psi}{\partial x}=0,
\label{main2}
\end{gather}
where $\Delta=\partial^{2}/\partial x^{2}+\partial^{2}/\partial
z^{2}$ is the two-dimensional Laplacian, and the Poisson bracket
(Jacobian) $\{f,g\}$ is defined by
\begin{equation}
\{f,g\}=\frac{\partial f}{\partial x}\frac{\partial g}{\partial z}
-\frac{\partial f}{\partial z}\frac{\partial g}{\partial x}.
\end{equation}
Here, $\psi(x,z)$ is the velocity stream function, $\chi (x,z)$ is
the normalized density perturbation, $H$ is the equivalent
atmospheric height, $\omega_{g}=(g/H)^{1/2}$ is the
Brunt-V\"{a}is\"{a}l\"{a}, $g$ is the free fall acceleration. The
coordinates $x$ and $z$ correspond to the horizontal and vertical
coordinates, respectively, and the $z$ axis is directed upward
against the gravitational acceleration
$\mathbf{g}=-g\hat{\mathbf{z}}$, and $\hat{\mathbf{z}}$ is the
unit vector along the $z$ direction and the $x$ axis lies in a
plane perpendicular to the $z$ axis. In the linear approximation,
taking $\psi\sim\exp (i\mathbf{K}\cdot\mathbf{x}-i\omega t)$ and
$\chi\sim\exp (i\mathbf{K}\cdot\mathbf{x}-i\omega t)$, where
$\mathbf{x}=(x,z)$, $\omega$ and $\mathbf{K}=(K_{x},K_{z})$ are
the frequency and wave number respectively, Eqs. (\ref{main1}) and
(\ref{main2}) yield the dispersion relation of the IGWs,
\begin{equation}
\label{dispers}
\omega^{2}=\frac{\omega_{g}^{2}K_{x}^{2}}{K^{2}+1/(4H^{2})},
\end{equation}
where $K^{2}=K_{x}^{2}+K_{z}^{2}$. In Eqs. (\ref{main1}) and
(\ref{main2}), characteristic frequencies $\omega\gg \Omega_{0}$
are considered, where $\Omega_{0}$ is the angular rotation
velocity of the planet, and then the Coriolis force can be
neglected. The Amp\`{e}re force is also neglected, that is, the
influence of the magnetic field in the ionized atmosphere, which
is justified at sufficiently high altitudes \cite{Kaladze2008}.
Further we consider an isothermal atmosphere, that is,
Brunt-V\"{a}is\"{a}l\"{a} frequency is assumed to be independent
of the vertical coordinate $z$. For the Earth's atmosphere, in
particular, this corresponds to altitudes $\gtrsim 200$ km. Due to
the dissipation of short-wave harmonics, the lower limit for
wavelengths for IGWs is $\sim 10$ km at altitudes $\sim 200$-$300$
km, while typical wavelength values are hundreds of kilometers.

The reductive perturbation method (the multiscale expansion
method)  \cite{Dodd1982} for Eqs. (\ref{main1}) and (\ref{main2})
to study the behavior of nonlinear IGWs was elaborated in a recent
paper \cite{Lashkin2024}. This method is often used in the theory
of nonlinear waves and leads to evolution equations, which in many
cases turn out to be more suitable for analysis than the original
problem. Following this technique, the space and time variables
were expanded in \cite{Lashkin2024} as
$\mathbf{x}=\mathbf{x}+\varepsilon\mathbf{X}+\dots$ and
$t=t+\varepsilon T+\varepsilon^{2}\tau+\dots$ respectively, where
$\mathbf{X}=(X,Z)$, and $\varepsilon$ is a small dimensionless
parameter that scales weak dispersion and nonlinearity. As was
shown, to obtain a non-trivial evolution, it was enough to
restrict ourselves to expanding the time variable up to the second
order and the space variable up to the first order in
$\varepsilon$, and in that case,
\begin{equation}
\label{exp-x-t} \frac{\partial}{\partial \mathbf{x}}\rightarrow
\frac{\partial}{\partial \mathbf{x}}+\varepsilon
\frac{\partial}{\partial \mathbf{X}}, \quad
\frac{\partial}{\partial t}\rightarrow \frac{\partial}{\partial
t}+\varepsilon \frac{\partial}{\partial T}+\varepsilon^{2}
\frac{\partial}{\partial \tau}.
\end{equation}
In turn, the fields $\psi$ and $\chi$ were expanded in powers in
$\varepsilon$ as
\begin{gather}
\psi=\varepsilon\psi^{(1)}+\varepsilon^{2}\psi^{(2)}+\varepsilon^{3}\psi^{(3)}+\dots,
\label{exp-psi}
\\
\chi=\varepsilon\chi^{(1)}+\varepsilon^{2}\chi^{(2)}+\varepsilon^{3}\chi^{(3)}+\dots,
\label{exp-phi}
\end{gather}
where $\psi^{(1)}=\tilde{\psi}^{(1)}+\bar{\psi}$,
$\chi^{(1)}=\tilde{\chi}^{(1)}+\bar{\chi}$,
\begin{gather}
\tilde{\psi}^{(1)}=\Psi
(\mathbf{X},T,\tau)\mathrm{e}^{i\mathbf{K}\cdot\mathbf{x}-i\omega
t}+\mathrm{c. c.}, \label{Psi}
\\
\tilde{\chi}^{(1)}=\Phi
(\mathbf{X},T,\tau)\mathrm{e}^{i\mathbf{K}\cdot\mathbf{x}-i\omega
t}+\mathrm{c. c.}. \label{Phi}
\end{gather}
Secondary mean flows $\bar{\psi}$ and $\bar{\chi}$ depend only on
slow variables $\mathbf{X}$, $T$ and $\tau$. As a result, using
the standard procedure for eliminating secular terms, the
following system of nonlinear equations for the envelope $\Psi$
and the secondary mean flow $\bar{\psi}$  was obtained in
\cite{Lashkin2024},
\begin{gather}
\hat{L}\Psi+K_{z}\left(K_{x}\frac{\omega_{g}^{2}}{v_{gx}}-\omega
K^{2}-\frac{K_{x}^{2}\omega_{g}^{2}}{\omega}\right)\Psi\frac{\partial\bar{\psi}}{\partial
X}
\nonumber \\
+K_{x}\left(\omega K^{2}
+\frac{K_{x}^{2}\omega_{g}^{2}}{\omega}\right)\Psi\frac{\partial\bar{\psi}}{\partial
Z}=0, \label{eq01}
\end{gather}
where
\begin{gather}
\hat{L}=2\omega\left(K^{2}+\frac{1}{4H^{2}}\right)\left(i\frac{\partial}{\partial
\tau}+\frac{1}{2}\frac{\partial^{2}\omega}{\partial
K_{x}^{2}}\frac{\partial^{2}}{\partial X^{2}}
+\frac{1}{2}\frac{\partial^{2}\omega}{\partial
K_{z}^{2}}\frac{\partial^{2}}{\partial Z^{2}} \right.
 \nonumber \\
\left. + \frac{\partial^{2}\omega}{\partial K_{x}\partial
K_{z}}\frac{\partial^{2}}{\partial X\partial Z}\right), \label{L2}
\end{gather}
\begin{gather}
\omega_{g}^{2}\frac{\partial^{2}\bar{\psi}}{\partial
X^{2}}-\frac{1}{4H^{2}}\left(v_{gx}^{2}\frac{\partial^{2}\bar{\psi}}{\partial
X^{2}}+v_{gz}^{2}\frac{\partial^{2}\bar{\psi}}{\partial
Z^{2}}\right) = \left(\omega
K^{2}+\frac{K_{x}^{2}\omega_{g}^{2}}{\omega}\right) \nonumber \\
\times\left(K_{z}\frac{\partial |\Psi|^{2}}{\partial
X}-K_{x}\frac{\partial |\Psi|^{2}}{\partial Z}\right),
\label{eq02}
\end{gather}
and for the group velocities $v_{gx}=\partial\omega/\partial
K_{x}$ and $v_{gz}=\partial\omega/\partial K_{z}$ we have
\begin{equation}
\label{vgx-vgz}
v_{gx}=\frac{\omega_{g}(K_{z}^{2}+1/4H^{2})}{(K^{2}+1/4H^{2})^{3/2}},
\quad v_{gz}=-\frac{\omega_{g}K_{x}K_{z}}{(K^{2}+1/4H^{2})^{3/2}}.
\end{equation}
In Ref.~\cite{Lashkin2024} only the one-dimensional case was
studied, when the system (\ref{eq01}) and (\ref{eq02}) was reduced
to either the focusing or defocusing one-dimensional NLS equation.
In the presented paper we consider the two-dimensional system.

We introduce dimensionless variables $\tau^{\prime}$, $x$, $z$,
$\Psi^{\prime}$ and $\bar{\psi}^{\prime}$  by
\begin{equation}
\label{dimensionless} \tau^{\prime}=\omega_{g}\tau, \,\,
x=\frac{X}{H}, \,\, z=\frac{Z}{H}, \,\,
\Psi^{\prime}=\frac{\Psi}{\omega_{g}H^{2}}, \,\,
\bar{\psi}^{\prime}=\frac{\bar{\psi}}{\omega_{g}H^{2}},
\end{equation}
and further the primes are omitted. Inserting Eq.
(\ref{dimensionless}) into Eqs. (\ref{eq01}) and (\ref{eq02})
yields
\begin{figure}
\includegraphics[width=3.in]{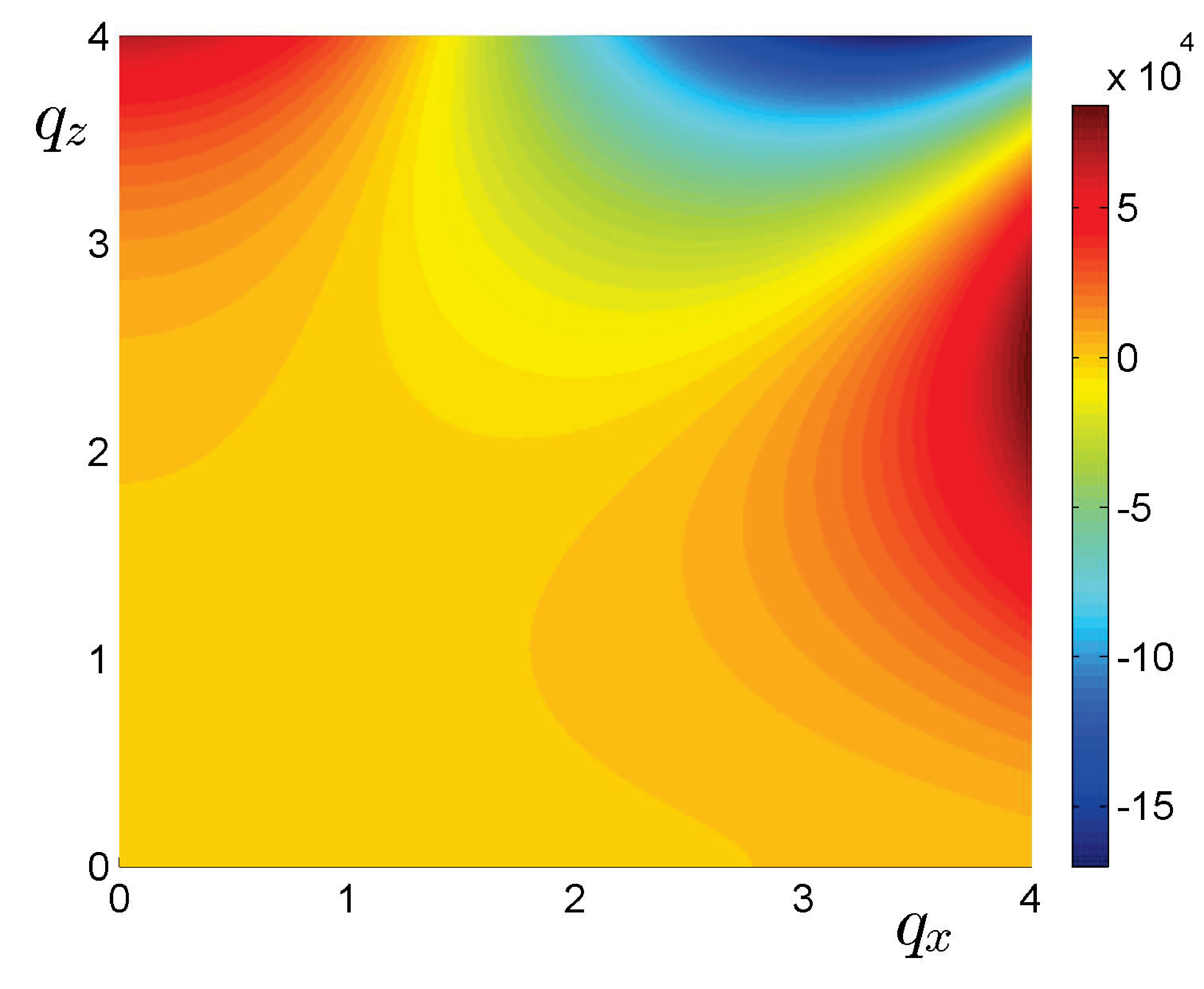}
\caption{\label{fig1} The contour plot of the function
$Q(q_{x},q_{z})$ in Eq. (\ref{Q}). Negative $Q$ corresponds to the
elliptic operator in the linear part of Eq. (\ref{eq1}).}
\end{figure}

\begin{figure}
\includegraphics[width=2.7in]{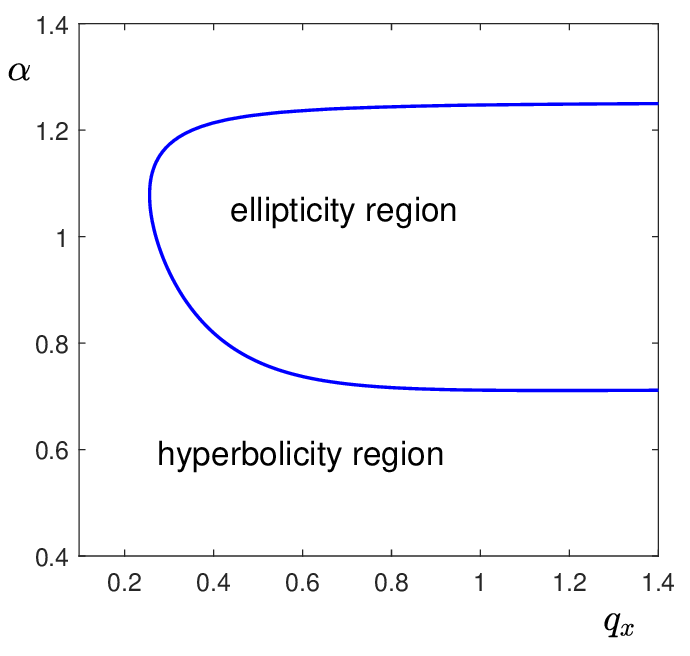}
\caption{\label{fig2} The regions of ellipticity and hyperbolicity
in Eq. (\ref{eq1}) on the plane $(q_{x},\alpha)$. These areas are
separated by the curve determined by the equation
$\tilde{Q}(q_{x},\alpha)=0$.}
\end{figure}

\begin{equation}
\label{eq1}
i\frac{\partial\Psi}{\partial\tau}+A\frac{\partial^{2}\Psi}{\partial
x^{2}}+B\frac{\partial^{2}\Psi}{\partial
z^{2}}+2C\frac{\partial^{2}\Psi}{\partial x \partial
z}+D\Psi\frac{\partial\bar{\psi}}{\partial
x}+E\Psi\frac{\partial\bar{\psi}}{\partial z}=0,
\end{equation}
\begin{equation}
\label{eq2} F\frac{\partial^{2}\bar{\psi}}{\partial
x^{2}}-G\frac{\partial^{2}\bar{\psi}}{\partial
z^{2}}=M\left(q_{z}\frac{\partial |\Psi|^{2}}{\partial
x}-q_{x}\frac{\partial |\Psi|^{2}}{\partial z}\right),
\end{equation}
where $q_{x}=K_{x}H$, $q_{z}=K_{z}H$, and the coefficients $A$,
$B$, $C$, $D$, $E$, $F$, $G$, and $M$ are determined in the
Appendix. Note that the coefficients $E$,$F$, $G$, and $M$ are
positive, $A$ is negative, while $B$, $C$, and $D$ are of
indefinite sign and their sign depends on the specific values of
$q_{x}$ and $q_{z}$. The properties of solutions to the system of
nonlinear equations (\ref{eq1}) and (\ref{eq2}) largely depend on
the linear part of Eq. (\ref{eq1}). The corresponding linear
partial differential equation with constant coefficients $A$, $B$,
and $C$ has elliptic type if
\begin{equation}
\label{elliptic} C^{2}-AB<0,
\end{equation}
and hyperbolic type if
\begin{equation}
\label{hyperbolic} C^{2}-AB>0.
\end{equation}
In case
\begin{equation}
\label{parabolic} C^{2}-AB=0,
\end{equation}
the equation is of parabolic type. Note that the classification of
equation types depends only on the coefficients of the second
derivatives and the sign of the first term in Eq. (\ref{eq1}) does
not affect it \cite{John1982}. As is known, the elliptic type of
equation corresponds to the boundary value problem, the hyperbolic
type to the wave equation, and the parabolic type to the diffusion
problem. In this paper, we restrict ourselves to the case
(\ref{elliptic}), that is, an elliptic operator in the linear part
of the equation (\ref{eq1}). The parabolic case (\ref{parabolic})
is not considered due to restrictions on the relationship between
$q_{x}$ and $q_{z}$, which greatly limits its practical
significance. The hyperbolic case (\ref{hyperbolic}) is expected
to be considered in the future. Using explicit expressions for
$A$, $B$ and $C$, conditions (\ref{elliptic}), (\ref{hyperbolic})
and (\ref{parabolic}) can be rewritten in equivalent forms $Q<0$,
$Q>0$ and $Q=0$, respectively, where the function $Q(q_{x},q_{z})$
is determined by
\begin{equation}
\label{Q} Q(q_{x},q_{z})=q_{z}^{2}(8q_{x}^{2}-4q_{z}^{2}-1)^{2}
-3q_{x}^{2}(1+4q_{z}^{2})(8q_{z}^{2}-4q_{x}^{2}-1).
\end{equation}
The contour plot of the function $Q(q_{x},q_{z})$ on the plane
$(q_{x},q_{z})$ is shown in Fig.~\ref{fig1}. By introducing the
angle $\alpha$ between the vector $\mathbf{q}=(q_{x},q_{z})$ and
the vector $q_{x}\hat{\mathbf{x}}$, where $\hat{\mathbf{x}}$ is a
unit vector in the horizontal direction, one can also define
another function $\tilde{Q}(q_{x},\alpha)$ with the same sign as
$Q(q_{x},q_{z})$ by
\begin{gather}
\tilde{Q}(q_{x},\alpha)=16q_{x}^{4}\tan^{2}\alpha
(\tan^{4}\alpha-10\tan^{2}\alpha+7) \nonumber \\
+4q_{x}^{2}(2\tan^{4}\alpha-7\tan^{2}\alpha+3)+3
 \label{Q1}
\end{gather}
The curve $\tilde{Q}(q_{x},\alpha)=0$ separating the plane
$(q_{x},\alpha)$ into the elliptic and hyperbolic regions of the
linear operator in Eq. (\ref{eq1}) is shown in Fig.~\ref{fig2}.
The ellipticity region of the linear operator in Eq. (\ref{eq1})
is visible from Fig.~\ref{fig1} and Fig.~\ref{fig2}. Note that
since $A<0$ it follows from (\ref{elliptic}) that $B<0$.

Using the convolution identity,
\begin{gather}
(fg)_{\mathbf{p},\omega}=\int
f_{\mathbf{p}_{1},\omega_{1}}g_{\mathbf{p}_{2},\omega_{2}}\delta
(\mathbf{p}-\mathbf{p}_{1}-\mathbf{p}_{2})\delta
(\omega-\omega_{1}-\omega_{2})
\nonumber \\
\times \, d\mathbf{p}_{1}d\mathbf{p}_{2}d\omega_{1}d\omega_{2},
\end{gather}
connecting the Fourier transforms of the product of arbitrary
functions $f(\mathbf{r},t)$ and $g(\mathbf{r},t)$ expressed in
physical space with the corresponding Fourier transforms of these
functions,
\begin{gather}
f_{\mathbf{p},\omega}=\int
f(\mathbf{r},t)\mathrm{e}^{-i\mathbf{p}\cdot\mathbf{r}+i\omega
t}d\mathbf{p}d\omega,
\\
g_{\mathbf{p},\omega}=\int
g(\mathbf{r},t)\mathrm{e}^{-i\mathbf{p}\cdot\mathbf{r}+i\omega
t}d\mathbf{p}d\omega,
\end{gather}
where $\delta (x)$ is the Dirac delta function, we can rewrite
Eqs. (\ref{eq1}) and (\ref{eq2}) in Fourier space as
\begin{equation}
\label{eq11} (\omega-\omega_{\mathbf{p}})\Psi_{p}\!=\!-i\!\!\int
\! (Dp_{2x}+Ep_{2z}) \Psi_{p_{1}}\bar{\psi}_{p_{2}}\delta
(p-p_{1}-p_{2})dp_{1}dp_{2},
\end{equation}
where
\begin{equation}
\omega_{\mathbf{p}}=Ap_{x}^{2}+Bp_{z}^{2}+2Cp_{x}p_{z},
\end{equation}
and
\begin{equation}
\label{eq22}
\bar{\psi}_{p}=\frac{iM(q_{z}p_{x}-q_{x}p_{z})}{Gp^{2}_{z}-Fp^{2}_{x}}\int
\Psi_{p_{1}}\Psi^{\ast}_{p_{2}}\delta (p-p_{1}-p_{2})dp_{1}dp_{2},
\end{equation}
respectively. Here and below we use the shorthand notation
$p\equiv(\mathbf{p},\omega)$, so that
\begin{equation}
\delta (p-p_{1}-p_{2})\equiv\delta
(\mathbf{p}-\mathbf{p}_{1}-\mathbf{p}_{2})\delta
(\omega-\omega_{1}-\omega_{2}),
\end{equation}
and $dp_{1}dp_{2}\equiv
d\mathbf{p}_{1}d\mathbf{p}_{2}d\omega_{1}d\omega_{2}$. Note that
from Eq. (\ref{elliptic}) it follows that $\omega_{\mathbf{p}}$ is
a negative definite quadratic form (despite the fact that the
coefficient $C$ is indefinite in sign), that is,
$\omega_{\mathbf{p}}<0$. Substituting Eq. (\ref{eq22}) into Eq.
(\ref{eq11}) we have one equation for the Fourier transform
$\Psi_{p}$,
\begin{equation}
\label{eq3} (\omega-\omega_{\mathbf{p}})\Psi_{p}\!=\!\int \!
V(p,p_{1},p_{2},p_{3})\Psi_{p_{1}}\Psi_{p_{2}}\Psi^{\ast}_{p_{3}}
dp_{1}dp_{2}dp_{3},
\end{equation}
where the interaction matrix element $V(p,p_{1},p_{2},p_{3})$ is
determined by
\begin{gather}
V(p,p_{1},p_{2},p_{3})=\frac{M}{2}\left[\frac{(Dp_{1x}+Ep_{1z})
(q_{z}p_{1x}-q_{x}p_{1z})}{Gp^{2}_{1z}-Fp^{2}_{1x}} \right.
\nonumber \\
\left. +
\frac{(Dp_{2x}+Ep_{2z})(q_{z}p_{2x}-q_{x}p_{2z})}{Gp^{2}_{2z}-Fp^{2}_{2x}}\right]
\delta (p-p_{1}-p_{2}-p_{3}), \label{V}
\end{gather}
and symmetrization in $p_{1}$ and $p_{1}$ is taken into account.
The system of nonlinear equations (\ref{eq1}) and (\ref{eq2}) has
apparently never been considered in problems in nonlinear physics
before. In appearance (and physical meaning), this system
resembles the Zakharov equations (and their generalizations)
describing the interaction of high-frequency and low-frequency
waves in plasma
\cite{Zakharov1972,Rubenchik1986,Lashkin2007POP,Lashkin2020} and
the equations for the interaction of short-wave and long-wave
disturbances on the surface of shallow water
\cite{Benney1977,Newell1978,Kanna2014}. The system (\ref{eq1}) and
(\ref{eq2}) is however much more difficult to analyze than the
analogues mentioned above.

Equation (\ref{eq3}) has an exact solution in the form of a
monochromatic plane wave,
\begin{equation}
\label{plane1} \Psi_{p}=\Psi_{0}V(p,p_{0},p_{0},-p_{0}),
\end{equation}
where
\begin{equation}
V(p,p_{0},p_{0},-p_{0})=S_{\mathbf{k}_{0}}|\Psi_{0}|^{2}\delta
(\mathbf{p}-\mathbf{k}_{0})\delta (\omega-\omega_{0}),
\end{equation}
and we have introduced the notation
\begin{equation}
\label{SK}
S_{\mathbf{k}_{0}}=\frac{M(Dk_{0x}+Ek_{0z})(q_{z}k_{0x}-q_{x}k_{0z})}{Gk_{0z}^{2}-Fk_{0x}^{2}}.
\end{equation}
In physical space this corresponds to the solution
\begin{equation}
\label{plane-wave} \Psi (\mathbf{r},t)=\Psi_{0} \exp
(i\mathbf{k}_{0}\cdot\mathbf{r}-i\omega_{0} t),
\end{equation}
with a frequency depending on the amplitude $\Psi_{0}$,
\begin{equation}
\label{omega0}
\omega_{0}=\omega_{\mathbf{k}_{0}}-S_{\mathbf{k}_{0}}|\Psi_{0}|^{2},
\end{equation}
where
\begin{equation}
\label{omega000}
\omega_{\mathbf{k}_{0}}=Ak_{0x}^{2}+Bk_{0z}^{2}+2Ck_{0x}k_{0z}.
\end{equation}
In the next section we consider the stability of such a plane
wave.

\section{\label {Sec3} Nonlinear dispersion relation and modulational instability}

The perturbed plane wave solution in physical space has the form
\begin{equation}
\label{pert1} \Psi=(\Psi_{0}+\delta\Psi)\exp
(i\mathbf{k}_{0}\cdot\mathbf{r}-i\omega_{0} t),
\end{equation}
where
\begin{equation}
\label{pert2}
\delta\Psi=\Psi^{+}\mathrm{e}^{i\mathbf{k}\cdot\mathbf{r}-i\Omega
t}+\Psi^{-}\mathrm{e}^{-i\mathbf{k}\cdot\mathbf{r}+i\Omega t},
\end{equation}
is a linear modulation with the frequency $\Omega$ and the wave
vector $\mathbf{k}$. In Fourier space, Eqs. (\ref{pert1}) and
(\ref{pert2}) correspond to
\begin{equation}
\label{pert11} \Psi_{p}=(\Psi_{0}+\delta\Psi_{p})\delta (p-p_{0}),
\end{equation}
and
\begin{equation}
\label{pert22} \delta\Psi_{p}=\Psi^{+}\delta
(\mathbf{p}-\mathbf{k})\delta (\omega-\Omega)+\Psi^{-}\delta
(\mathbf{p}+\mathbf{k})\delta (\omega+\Omega),
\end{equation}
respectively. Linearizing Eq. (\ref{eq3}) in $\delta\Psi_{p}$, we
get the nonlinear dispersion relation,
\begin{equation}
\label{nonlin-disp} 1-|\Psi_{0}|^{2}
\left[\frac{S_{\mathbf{k}_{0}+\mathbf{k}}}{\omega_{\mathbf{k}_{0}
+\mathbf{k}}-\omega_{\mathbf{k}_{0}}-\Omega}
+\frac{S_{\mathbf{k}_{0}-\mathbf{k}}}
{\omega_{\mathbf{k}_{0}-\mathbf{k}}-\omega_{\mathbf{k}_{0}}+\Omega}\right]=0.
\end{equation}
Equation (\ref{nonlin-disp}) is a quadratic equation in $\Omega$,
\begin{gather}
\Omega^{2}+\Omega[\omega_{\mathbf{k}_{0}
-\mathbf{k}}-\omega_{\mathbf{k}_{0} +\mathbf{k}}
+(S_{\mathbf{k}_{0}+\mathbf{k}}-S_{\mathbf{k}_{0}-\mathbf{k}})|\Psi_{0}|^{2}]
\nonumber \\
+[S_{\mathbf{k}_{0}+\mathbf{k}}
(\omega_{\mathbf{k}_{0}-\mathbf{k}}-\omega_{\mathbf{k}_{0}})
+S_{\mathbf{k}_{0}-\mathbf{k}}
(\omega_{\mathbf{k}_{0}+\mathbf{k}}-\omega_{\mathbf{k}_{0}})]|\Psi_{0}|^{2}
\nonumber \\
-(\omega_{\mathbf{k}_{0}+\mathbf{k}}-\omega_{\mathbf{k}_{0}})
(\omega_{\mathbf{k}_{0}-\mathbf{k}}- \omega_{\mathbf{k}_{0}})=0,
\label{nonlin-disp1}
\end{gather}
and it can be easily solved. The negativity of the discriminant of
the equation corresponds to instability with the growth rate
$\gamma=|\mathrm{Im}\,\Omega|$. It can also be seen that the
instability has a threshold character with respect to the
amplitude $\Psi_{0}$. As noted above, the coefficients $C$ and $D$
are indefinite in sign and their sign depends on the specific
values of $q_{x}$ and $q_{z}$. In the general case, the dependence
of the instability growth rate on the wave vector of a plane wave
$\mathbf{k}_{0}$, the wave vector of perturbations $\mathbf{k}$,
and the values of $q_{x}$ and $q_{z}$ is quite complex. Equation
(\ref{nonlin-disp1}) is greatly simplified in a number of
important limiting cases. In the limit of long-wave modulations
$\mathbf{k}\ll\mathbf{k}_{0}$, using
\begin{equation}
\omega_{\mathbf{k}_{0}\pm\mathbf{k}}\sim
\omega_{\mathbf{k}_{0}}\pm \frac{\partial
\omega_{\mathbf{k}_{0}}}{\partial
\mathbf{k}_{0}}\cdot\mathbf{k}+\frac{1}{2}\frac{\partial^{2}
\omega_{\mathbf{k}_{0}}}{\partial \mathbf{k}_{0}^{2}}k^{2},
\end{equation}
from Eq. (\ref{nonlin-disp1}) we obtain
\begin{equation}
\label{nonlin-disp-convective} (\Omega-\mathbf{v}_{g}\cdot
\mathbf{k})^{2}=\frac{1}{4}(\omega_{\mathbf{k}_{0}}^{\prime\prime})^{2}k^{4}
-\omega_{\mathbf{k}_{0}}^{\prime\prime}k^{2}S_{\mathbf{k}_{0}}|\Psi_{0}|^{2},
\end{equation}
where $\mathbf{v}_{g}=\partial \omega_{\mathbf{k}_{0}}/\partial
\mathbf{k}_{0}$ and
$\omega_{\mathbf{k}_{0}}^{\prime\prime}=\partial^{2}
\omega_{\mathbf{k}_{0}}/\partial \mathbf{k}_{0}^{2}$. Since
$\omega_{\mathbf{k}_{0}}<0$, then if $S_{\mathbf{k}_{0}}<0$ and
the amplitude threshold is exceeded,
\begin{equation}
4|S_{\mathbf{k}_{0}}||\Psi_{0}|^{2}>|\omega_{\mathbf{k}_{0}}^{\prime\prime}|k^{2},
\end{equation}
equation (\ref{nonlin-disp-convective}) corresponds to convective
instability when growing disturbances are carried away with the
group velocity $\mathbf{v}_{g}$, and the instability growth rate
is given by
\begin{equation}
\gamma=k\sqrt{4|S_{\mathbf{k}_{0}}||\omega_{\mathbf{k}_{0}}^{\prime\prime}|
|\Psi_{0}|^{2}-(\omega_{\mathbf{k}_{0}}^{\prime\prime})^{2}k^{2}}.
\end{equation}
Note that in this case the instability with respect to the wave
numbers of perturbations $k_{x}$ and $k_{z}$ is isotropic.

More interesting is the opposite case of short-wave modulations
$\mathbf{k}\gg\mathbf{k}_{0}$. This instability is an instability
of a uniform field (in the limit $\mathbf{k}_{0}\rightarrow 0$)
leading to the splitting of this field into clumps, which
ultimately results in the emergence of coherent structures at the
nonlinear stage, which generally speaking can be both
non-stationary (collapsing cavitons) \cite{Zakharov1972} and
stationary (stable solitons). Then, taking into account that
$\omega_{\mathbf{k}}$ and $S_{\mathbf{k}}$ are even functions, Eq.
(\ref{nonlin-disp1}) becomes
\begin{equation}
\label{nonlin-disp-pure}
\Omega^{2}=\omega_{\mathbf{k}}\left(\omega_{\mathbf{k}}-2S_{\mathbf{k}}|\Psi_{0}|^{2}\right).
\end{equation}
Since $\omega_{\mathbf{k}}<0$, then Eq. (\ref{nonlin-disp-pure})
predicts a purely growing instability (modulational instability)
if $S_{\mathbf{k}}<0$ and if the amplitude threshold is exceeded,
\begin{equation}
 2|S_{\mathbf{k}}||\Psi_{0}|^{2}>|\omega_{\mathbf{k}}|.
\end{equation}
The instability growth rate $\gamma$ is given by
\begin{equation}
\label{pure}
\gamma=\sqrt{2|S_{\mathbf{k}}||\omega_{\mathbf{k}}||\Psi_{0}|^{2}-\omega_{\mathbf{k}}^{2}}.
\end{equation}
From Eqs. (\ref{SK}) and (\ref{omega000}) in which
$\mathbf{k}_{0}$ is replaced by $\mathbf{k}$ it is evident that
instability has a substantially anisotropic character and,
depending on the relationship between $k_{x}$ and $k_{z}$, the
mode of modulation instability can change (longitudinal or
transverse instability of the envelope wave packets). Moreover, it
follows from Eq. (\ref{SK}) that the parametric coupling is
nonlocal. By introducing the angle $\beta$ between the vector
$\mathbf{k} = (k_{x}, k_{z})$ and the vector
$k_{x}\hat{\mathbf{x}}$, where $\hat{\mathbf{x}}$ is a unit vector
in the $x$-direction, one can rewrite $S_{\mathbf{k}}$ in Eq.
(\ref{SK}) as
\begin{equation}
\label{Sk-1}
 S_{\mathbf{k}}=\frac{M(D+E\tan\beta)(q_{z}-q_{x}\tan\beta)}{F(\varepsilon\tan^{2}\beta-1)},
\end{equation}
where $\varepsilon=G/F$ and for typical values of carrier wave
numbers $q_{x}$ and $q_{z}$ we have $\varepsilon\ll 1$. Equation
(\ref{Sk-1}) is not valid for $\varepsilon\tan^{2}\beta\sim 1$ (in
this case, the conditions for deriving Eq. (\ref{eq2}) are
violated). First we consider the angles $\beta$ satisfying
condition $\varepsilon\tan^{2}\beta \ll 1$. Note that this case
corresponds to both longitudinal $k_{x}\gg k_{z}$ and transverse
$k_{z}\gg k_{x}$ modulation instability, except for very small
longitudinal wave numbers $k_{x}$ corresponding to angles
$\beta\gg\arctan (\sqrt{1/\varepsilon})$.
\begin{figure}
\includegraphics[width=3.in]{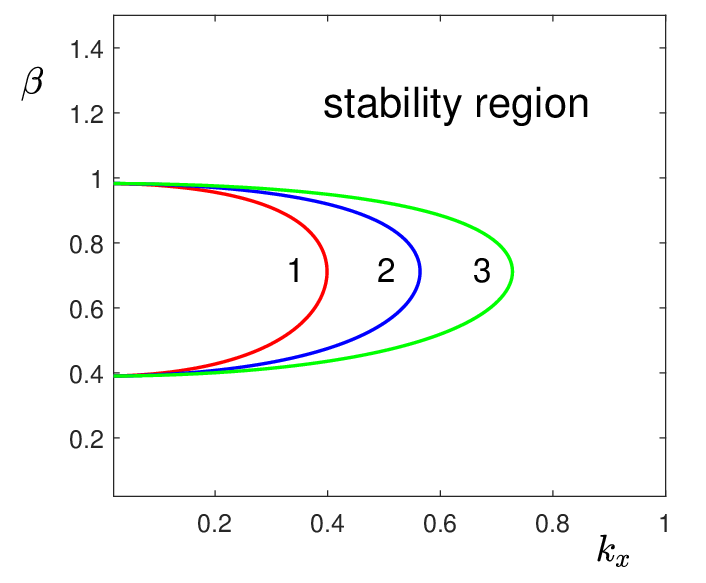}
\caption{\label{fig3} Stability and instability regions on the
plane $(k_{x},\beta)$ for different amplitude values $\Psi_{0}$.
The outer regions to the right of the curves correspond to the
stability regions. The numbers near the curves correspond to
different amplitudes: 1 - $\Psi_{0}=0.003$, 2 - $\Psi_{0}=0.006$,
and 3 - $\Psi_{0}=0.01$.}
\end{figure}
The stability region with respect to the wave numbers of
perturbations and the amplitude of the plane wave depends on the
sign of the function $\mathcal{Q}(k_{x},\beta)$ defined by
\begin{gather}
\mathcal{Q}(k_{x},\beta)=2|\Psi_{0}|^{2}M(D+E\tan\beta)(q_{x}\tan\beta-q_{z})/F
\nonumber \\
+k_{x}^{2}(A+B\tan^{2}\beta+2C\tan\beta).  \label{Om-critic}
\end{gather}
The curves $\mathcal{Q}(k_{x},\beta)=0$, dividing the plane
$(k_{x},\beta)$ into stable and unstable regions for different
values of the plane wave amplitude $\Psi_{0}$ and for specific
values of the carrier wave numbers $q_{x}=1$ and $q_{z}=1.5$ are
shown in Fig.~\ref{fig3}. The picture does not qualitatively
depend on the specific values of the carrier wave numbers $q_{x}$
and $q_{z}$ (in the considered region of ellipticity) and the
amplitude $\Psi_{0}$. It is evident from Fig.~\ref{fig3} that the
instability is anisotropic, and a mode of change of modulation
instability from longitudinal to transverse is possible. In the
case $\varepsilon\tan^{2}\beta \gg 1$, the modulation instability
is only transverse.
\begin{figure}
\includegraphics[width=3.in]{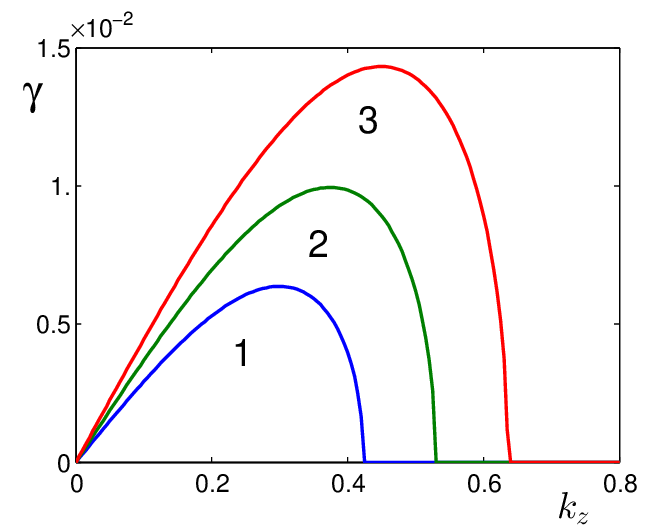}
\caption{\label{fig4} Dependence of the instability growth rate
$\gamma$ on the vertical wave number $k_{z}$ in the case $k_{z}\gg
k_{x}$ for different amplitude values $\Psi_{0}$. The numbers
under the curves correspond to different amplitudes: 1 -
$\Psi_{0}=0.004$, 2 - $\Psi_{0}=0.005$, and 3 - $\Psi_{0}=0.006$.}
\end{figure}
It can be seen that $S_{\mathbf{k}}$ does not depend on
$\mathbf{k}$ for perturbations with $k_{z}\ll k_{x}$ (pure
longitudinal instability) or $k_{z}\gg k_{x}$ (pure transverse
instability). In these cases, the expression for $S_{\mathbf{k}}$
is reduced to $S_{1}$ and $S_{2}$ defined as
\begin{equation}
\label{S1S2} S_{1}=-\frac{MDq_{z}}{F}, \quad
S_{2}=-\frac{MEq_{x}}{G}.
\end{equation}
In the ellipticity region we have $D>0$, and all other
coefficients in Eq.~(\ref{S1S2}) are always positive, so that the
necessary conditions for instability $S_{1}<0$ and $S_{2}<0$ are
satisfied. Then the longitudinal and transverse instability growth
rates are obtained from Eq. (\ref{pure}), and have the form,
respectively,
\begin{equation}
\label{g1}
\gamma=\sqrt{2S_{1}Ak_{x}^{2}|\Psi_{0}|^{2}-A^{2}k_{x}^{4}}, \quad
\mathrm{if} \,\, k_{z}\ll k_{x},
\end{equation}
and
\begin{equation}
\label{g2}
\gamma=\sqrt{2S_{2}Bk_{z}^{2}|\Psi_{0}|^{2}-B^{2}k_{z}^{4}}, \quad
\mathrm{if} \,\, k_{z}\gg k_{x}.
\end{equation}
The optimal horizontal and vertical wave numbers of perturbations
corresponding to the maximum instability growth rates in
(\ref{g1}) and (\ref{g2}) are
\begin{equation}
\label{k-opt}
k_{x,\mathrm{opt}}=|\Psi_{0}|\sqrt{\frac{S_{1}}{|A|}}  \quad
\mathrm{and}  \quad
k_{z,\mathrm{opt}}=|\Psi_{0}|\sqrt{\frac{S_{2}}{|B|}},
\end{equation}
respectively. It is at such scales that instability most
contributes to the emergence of coherent nonlinear entities
(stationary or nonstationary). In fact, just on such scales a
two-dimensional soliton (which apparently turns out to be
unstable) or a collapsing caviton can arise. The dependence of the
instability growth rate $\gamma$ on the vertical wave number of
perturbations $k_{z}$ in the case $k_{z}\gg k_{x}$ for different
amplitude values $\Psi_{0}$ and for specific values $q_{x}=1$ and
$q_{z}=1.5$ is shown in Fig.~\ref{fig4}. For example, in the
Earth's atmosphere, the equivalent atmospheric height at the
altitudes $\gtrsim 200$ km (i.e. for an isothermal atmosphere) is
$H\sim 40$ km, that is the values of $q_{x}$ and $q_{z}$
correspond to horizontal and vertical wavelengths $\sim 40$ km and
$\sim 30$ km, respectively. Note that the wave numbers of
perturbation (envelope) are much less than the characteristic wave
numbers of the IGW (carrier). For the value $\Psi_{0}=0.004$,
which corresponds to the perturbation velocity $\sim 10$ m/s, the
optimal values $k_{z,\mathrm{opt}}=0.3$ corresponds to the
characteristic vertical size of the perturbation region $\sim 130$
km.

\section{\label {Sec4} Collapse of internal gravity waves}

The system of Eqs. (\ref{eq1}) and (\ref{eq2}) can be written in
the form of the generalized nonlinear Schr\"{o}dinger equation
with the nonlocal nonlinearity,
\begin{gather}
i\frac{\partial\Psi}{\partial\tau}+A\frac{\partial^{2}\Psi}{\partial
x^{2}}+B\frac{\partial^{2}\Psi}{\partial
z^{2}}+2C\frac{\partial^{2}\Psi}{\partial x \partial z} \nonumber
\\
+\Psi\int R(\mathbf{r}-\mathbf{r}')M\left(q_{z}\frac{\partial
}{\partial x}-q_{x}\frac{\partial}{\partial
z}\right)|\Psi(\mathbf{r}')|^{2} d^{2}\mathbf{r}'=0, \label{eq4}
\end{gather}
where the kernel $R(\mathbf{r})$ is the Green's function of the
equation
\begin{equation}
\label{Green} \left(F\frac{\partial^{2}}{\partial
x^{2}}-G\frac{\partial^{2}}{\partial
z^{2}}\right)R(\mathbf{r}-\mathbf{r}')=\delta(\mathbf{r}-\mathbf{r}^{'}),
\end{equation}
that in Fourier space corresponds to
\begin{equation}
\label{Greenk} R_{\mathbf{k}}=\frac{1}{Gk_{z}^{2}-Fk_{x}^{2}}.
\end{equation}
Equation (\ref{eq4}) conserves the 2D norm
\begin{equation}
\label{N} \mathcal{N}=\int |\Psi|^{2}d^{2}\mathbf{r},
\end{equation}
and Hamiltonian
\begin{gather}
\mathcal{H}=\int \left\{A\left|\frac{\partial\Psi}{\partial
x}\right|^{2}+B\left|\frac{\partial\Psi}{\partial
z}\right|^{2}+\frac{C}{2}\left(\frac{\partial\Psi}{\partial
x}\frac{\partial\Psi^{\ast}}{\partial
z}+\frac{\partial\Psi}{\partial
z}\frac{\partial\Psi^{\ast}}{\partial x}\right)
 \right. \nonumber
\\
\left. -\frac{|\Psi|^{2}}{2}\int
R(\mathbf{r}-\mathbf{r}')M\left(q_{z}\frac{\partial }{\partial
x}-q_{x}\frac{\partial}{\partial z}\right)|\Psi(\mathbf{r}')|^{2}
d^{2}\mathbf{r}'\right\}d^{2}\mathbf{r},
\end{gather}
and can be written in the hamiltonian form
\begin{equation}
i\frac{\partial\Psi}{\partial
t}=\frac{\delta\mathcal{H}}{\delta\Psi^{\ast}}.
\end{equation}
The nonlinear term in Eq. (\ref{eq4}) is somewhat reminiscent of
the nonlocal nonlinearity in previously studied models with a
kernel depending on the difference in spatial coordinates. The
expressions for the kernels $R(\mathbf{r})$ in these models were
dictated by the corresponding physical problems and were quite
different from each other. For example, in Ref.~\cite{Lashkin2006}
the kernel has the form of a quadratic exponential, in
Ref.~\cite{Lashkin2007PLA} a Hankel function of the first kind of
zero order, and in Ref.~\cite{Lashkin2007PRA} it contains the
complementary error function. In all these cases, the nonlocal
nonlinearity was scale inhomogeneous in spatial variables, and
this is what resulted in the absence of collapse and the existence
of stable coherent structures in the form of not only the 2D
fundamental soliton (the ground state), but also in the form of a
dipole soliton, rotating multisolitons and vortex solitons.

The key point for further analysis is the scale homogeneity of Eq.
(\ref{eq4}) in spatial variables, which is easily seen from Eq.
(\ref{Greenk}), and as a consequence, the Hamiltonian
$\mathcal{H}$. The stationary solution of Eq. (\ref{eq4}) in the
form of $\Psi (\mathbf{r},t)=\Phi (\mathbf{r})\exp
(i\lambda^{2}t)$ corresponds to a stationary point of the
Hamiltonian $\mathcal{H}$ for a fixed 2D norm $\mathcal{N}$ and
resolves the variational problem $\delta\mathcal{S}[\Phi]=0$ for
the functional
\begin{equation}
\label{S-func}
\mathcal{S}[\Phi]=\mathcal{H}+\lambda^{2}\mathcal{N}.
\end{equation}
Solving this variational problem is equivalent to finding a
solution of the stationary equation
\begin{gather}
-\lambda^{2}\Phi+A\frac{\partial^{2}\Phi}{\partial
x^{2}}+B\frac{\partial^{2}\Phi}{\partial
z^{2}}+2C\frac{\partial^{2}\Phi}{\partial x \partial z} \nonumber
\\
+\Phi\int R(\mathbf{r}-\mathbf{r}')M\left(q_{z}\frac{\partial
}{\partial x}-q_{x}\frac{\partial}{\partial
z}\right)|\Phi(\mathbf{r}')|^{2} d^{2}\mathbf{r}'=0. \label{eq5}
\end{gather}
Multiplying Eq. (\ref{eq5}) by $\Phi^{\ast}$, and then integrating
over the whole space (taking into account zero boundary conditions
at infinity) we obtain,
\begin{equation}
\label{scale1} -\lambda^{2}\mathcal{N}-I_{1}+I_{2}=0,
\end{equation}
where
\begin{equation}
\label{I1}   I_{1}=A\left|\frac{\partial\Phi}{\partial
x}\right|^{2}+B\left|\frac{\partial\Phi}{\partial
z}\right|^{2}+\frac{C}{2}\left(\frac{\partial\Phi}{\partial
x}\frac{\partial\Phi^{\ast}}{\partial
z}+\frac{\partial\Phi}{\partial
z}\frac{\partial\Phi^{\ast}}{\partial x}\right),
\end{equation}
and
\begin{equation}
\label{I2}   I_{2}=|\Phi|^{2}\int
R(\mathbf{r}-\mathbf{r}')M\left(q_{z}\frac{\partial }{\partial
x}-q_{x}\frac{\partial}{\partial z}\right)|\Phi(\mathbf{r}')|^{2}
d^{2}\mathbf{r}'d^{2}\mathbf{r}.
\end{equation}
On the other hand, one can write
\begin{equation}
\label{scale2} \mathcal{H}=I_{1}-\frac{I_{2}}{2}.
\end{equation}
With the scale homogeneity in mind, we consider an
$\mathcal{N}$-preserving scaling transformation
$\Phi^{(\alpha)}=\Phi (\alpha\mathbf{r})$ and obtain for the
corresponding values,
\begin{equation}
\label{alpha} \mathcal{N}^{(\alpha)}=\alpha^{2}\mathcal{N}, \quad
I_{1}^{(\alpha)}=I_{1}, \quad I_{2}^{(\alpha)}=\alpha^{2}I_{2}.
\end{equation}
We use the approach developed by Hobart and Derrick
\cite{Hobart1963,Derrick1964} (see also further development in
Ref.~\cite{Makhankov1993}). For the functional $V[\phi]$ of the
form
\begin{equation}
V[\phi]=\sum_{\nu=-n_{1}}^{n_{2}}V^{(\nu)}(\alpha)
\end{equation}
where $V^{(\nu)}(\alpha)$ is a homogeneous function of the scale
parameter $\alpha$ of degree $\nu$, and with a stationary point
$\phi=u(\mathbf{r})$, that is $\delta V[u]=0$, with a scale
transformation $\phi_{\alpha}=u(\alpha\mathbf{r})$, the following
equality
\begin{equation}
\frac{\delta
V[u]}{\delta\alpha}=\sum_{\nu=-n_{1}}^{n_{2}}\left.\frac{\partial
V^{(\nu)}}{\partial\alpha}\right|_{\alpha=1}=\sum_{\nu=-n_{1}}^{n_{2}}\left.\nu
V^{(\nu)}\right|_{\alpha=1}=0
\end{equation}
is true (the so called virial theorem). Since the functional
$\mathcal{S}$ is scale homogeneous in spatial variables, from the
Hobart-Derrick virial theorem we can immediately write,
\begin{equation}
\label{Derric} \left.\frac{\partial}{\partial
\alpha}(\mathcal{H}^{(\alpha)}+\lambda^{2}\mathcal{N}^{(\alpha)})\right|_{\alpha=1}=0,
\end{equation}
and then from Eqs. (\ref{alpha}) and (\ref{Derric}) we have an
additional restriction for the stationary states,
\begin{equation}
\label{scale3}  2\lambda^{2}\mathcal{N}+I_{2}=0.
\end{equation}
Combining Eqs. (\ref{scale1}), (\ref{scale2}), and (\ref{scale3})
one can obtain that $\mathcal{H}=0$. Thus, the Hamiltonian at any
stationary solution is equal to zero. This fact in the model under
consideration is not accidental. This is typical for the 2D models
with a cubic local nonlinearity and scale homogeneity in both
spatial variables. Then, one can conclude that an arbitrary
initial localized field distribution with $\mathcal{H}\neq 0$
never reaches a stationary state in the course of evolution, that
is, either spreads out or collapses. Collapse in the model of the
two-dimensional NLS equation is usually called critical, since
(unlike the three-dimensional case) it occurs when the 2D norm of
the wave field exceeds a certain critical value (in this case the
Hamiltonian is negative) \cite{Sulem1999}. The same is true for
our model. By analogy with the two-dimensional NLS equation, one
can expect for the critical value $\mathcal{N}_{c}=\int
\Phi_{0}^{2}d^{2}\mathbf{r}$, where $\Phi_{0}(\mathbf{r})$ is a
nodeless solution (ground state) of Eq. (\ref{eq5}). Thus, a
sufficiently intense disturbance results in the collapse of
internal gravity waves.

\section{\label {Sec5} Conclusion}

We have studied the dynamics of 2D nonlinear IGWs. The analysis
was carried out on the basis of a system of 2D nonlinear equations
for the velocity stream function and secondary mean flow, obtained
with the aid of the reductive perturbation method in
Ref.~\cite{Lashkin2024}. We have obtained one equation for the
envelope in the form of 2D generalized nonlinear Schr\"{o}dinger
equation with nonlocal nonlinearity when the nonlinear response
depends on the wave intensity at some spatial domain. Only the
elliptic type of this equation has been considered. The
instability of a monochromatic plane wave, which is an exact
solution of the corresponding equation, has been studied, and a
nonlinear dispersion equation has been found. In the limit of
long-wave modulations, when the wave vector of modulations can be
neglected compared to the wave vector of the plane wave, the
instability is of a convective type. In the opposite case of
short-wave modulations, we have a purely growing instability
(modulation instability). In both cases, the corresponding
instability thresholds and instability growth rates have been
found. Numerical estimates for a characteristic region of
localization of unstable perturbations, consistent with the
results of possible predictions of experimental observations on
nonlinear IGWs, are given for the real Earth's atmosphere. It is
usually believed that modulation instability at the nonlinear
stage results in the formation of a soliton or collapsing caviton.
We have shown that, due to scaling homogeneity in spatial
variables, the Hamiltonian of the resulting nonlinear equation
with nonlocal nonlinearity is equal to zero, which leads to the
critical collapse of atmospheric IGWs. In reality, no singularity
occurs and the collapse arrests due to the dissipation of
short-wave harmonics corresponding to the lower limit of
wavelengths for IGWs ($\lesssim 10$ km for the Earth's atmosphere
at altitudes $\sim 200$ km).

\appendix
\section{}

In this appendix we write down the explicit expressions for the
coefficients $A$, $B$, $C$, $D$, $E$, $F$, $G$, and $M$ in Eqs.
(\ref{eq1}) and (\ref{eq2}),
\begin{gather}
A=-\frac{12q_{x}(1+4q_{z}^{2})}{(1+4q^{2})^{5/2}}, \quad
B=-\frac{4q_{x}(8q_{z}^{2}-4q_{x}^{2}-1)}{(1+4q^{2})^{5/2}}, \\
C=\frac{4q_{z}(8q_{x}^{2}-4q_{z}^{2}-1)}{(1+4q^{2})^{5/2}}, \quad
D=\frac{2q_{z}(4q_{x}^{4}-4q_{z}^{4}-q_{z}^{2})}{(1+4q_{z}^{2})(1+4q^{2})},
\\ E=\frac{q_{x}(1+8q^{2})}{2(1+4q^{2})},\quad
F=\left[1-\frac{(1+4q_{z}^{2})^{2}}{(1+4q^{2})^{3}}\right] ,
\\
G=\frac{16q_{x}^{2}q_{z}^{2}}{(1+4q^{2})^{3}}, \quad
M=\frac{q_{x}(1+8q^{2})}{2(1+4q^{2})^{1/2}}.
\end{gather}


\begin{thebibliography}{59}

\bibitem{Hines1960}
C.~O. Hines, Internal atmospheric gravity waves at ionospheric
heights, Can. J.  Phys. \textbf{38}, 1441-1481 (1960).

\bibitem{Tolstoy1967}
I.~Tolstoy, Long-period gravity waves in the atmosphere, J.
Geophys. Res. \textbf{72},  4605-4610 (1967).

\bibitem{Waltercheid2005}
R. L. Waltercheid, J. H. Hecht,  A reexamination of evanescent
acoustic-gravity waves: Special properties and aeronomical
significance, J. Geophys. Res. \textbf{108} (D11) 4340 (2005)

\bibitem{Cheremnykh2019}
O. K. Cheremnykh, A. K. Fedorenko, E. I. Kryuchkov, Y. A.
Selivanov, Evanescent acoustic-gravity modes in the isothermal
atmosphere: systematization, applications to the Earth's and Solar
atmospheres, Ann. Geophys. \textbf{37} 405-415 (2019).

\bibitem{Cheremnykh2021}
O. Cheremnykh, A. Fedorenko, Y. Selivanov, S. Cheremnykh,
Continuous spectrum of evanescent acoustic-gravity waves in an
isothermal atmosphere, Month. Notic. of the Royal Astronomical
Society \textbf{503},  5545-5553 (2021).

\bibitem{Liu1974}
K.~C. Yeh, C.~H. Liu, Acoustic-gravity waves in the upper
atmosphere, Rev.  Geophys. Space Phys. \textbf{12}, 193-216
(1974).

\bibitem{Beer1974}
T.~Beer, \emph{Atmospheric Waves}  (John Wiley, New York, 1974).

\bibitem{Gossard1975}
E.~E. Gossard, W.~H. Hooke, \emph{Waves in the Atmosphere:
Atmospheric Infrasound and  Gravity Waves: Their Generation and
Propagation}  (Elsevier Scientific Publishing Company, 1975).

\bibitem{Schubert1984}
G. Schubert,  R. L. Walterscheid, Propagation of small-scale
acoustic-gravity waves in the Venus atmosphere, Journal of the
Atmospheric Sciences, \textbf{41}, 1202-1213 (1984).

\bibitem{Forbes2009}
J. M. Forbes, Y. Moudden, Solar terminator wave in a Mars general
circulation model, Geophys. Res. Lett. \textbf{36} 17201 (2009).

\bibitem{Prikryl2005}
P. Prikryl, D. B. Muldren, S. J. Sofko, J. M. Ruohoniemi, Solar
wind Alfven waves: a source of pulsed ionospheric convection and
atmospheric gravity waves, Ann. Geophys. \textbf{23}, 401-417
(2005).

\bibitem{Fritts2008}
D. C. Fritts, S. L. Vadas, Gravity wave penetration into the
thermosphere: sensitivity to solar cycle variations and mean
winds,  Ann. Geophys. \textbf{26}, 3841-3861 (2008).

\bibitem{Bespalova2016}
A. V. Bespalova, A. K. Fedorenko, O. K. Cheremnykh, I. T. Zhuk,
Satellite observations of wave disturbances caused by moving solar
terminator, J. Atmos. Solar. Terr. Phys.  \textbf{140}, 79-85
(2016).

\bibitem{Rapoport2004}
Yu. G. Rapoport, O. E. Gotynyan, V. M. Ivchenko, L. V. Kozak, M.
Parrot, Effect of acoustic-gravity wave of the lithospheric origin
on the ionospheric F region before earthquakes, Phys. and Chem.
Earth \textbf{29}, 607-616 (2004).

\bibitem{Yigit2008}
E. Yi\u{g}it, A. D. Aylward, A. S. Medvedev, Parameterization of
the effects of vertically propagating gravity waves for
thermosphere general circulation models: Sensitivity study, J.
Geophys. Res.
 \textbf{113}, D19106 (2008).

\bibitem{Sutherland2015}
B. R. Sutherland, \emph{Internal Gravity Waves}  (Cambridge
University Press, Cambridge, 2015).

\bibitem{Francis1975}
S. H. Francis, Global propagation of atmospheric gravity waves: A
review, J. Atmos. Sol.-Terrestrial Phys. \textbf{37}, 1011-1054
(1975).

\bibitem{Fritts2003}
D. C. Fritts and M. J. Alexander, Gravity wave dynamics and
effects in the middle atmosphere, Rev. Geophys. \textbf{41},
1003-1062 (2003).

\bibitem{Kaladze2008}
T. D. Kaladze, O.~A. Pokhotelov, H.~A. Shah, M.~I. Khan, L.
Stenflo, Acoustic-gravity waves in the Earth's ionosphere, J.
Atmos. Sol.-Terrestrial Phys. \textbf{70}, 1607-1616 (2008).

\bibitem{Lashkin2023}
V.~M. Lashkin and O.~K. Cheremnykh, Acoustic-gravity waves in
quasi-isothermal atmospheres with a random vertical temperature
profile, Wave Motion \textbf{119}, 103140 (2023).

\bibitem{Roy2019}
A. Roy, S. Roy, A.~P. Misra, Dynamical properties of
acoustic-gravity waves in the atmosphere, J. of Atmos. and
Solar-Terr. Phys. \textbf{186}, 78-81 (2019).

\bibitem{Miropolsky2001}
Yu. Z. Miropol'sky, \emph{Dynamics of Internal Gravity Waves in
the Ocean}  (Kluwer, Dordrecht, 2001).

\bibitem{Stenflo1987}
L.~Stenflo, Acoustic solitary waves, Phys. Fluids \textbf{30},
3297-3299 (1987).

\bibitem{Stenflo1990}
L.~Stenflo, Acoustic gravity vortices, Phys. Scripta \textbf{41},
641-642 (1990).

\bibitem{Stenflo2009}
L.~Stenflo and P.~K. Shukla, Nonlinear acoustic-gravity waves, J.
Plasma Physics \textbf{75}, 841-847 (2009).

\bibitem{Dong1988}
B. Dong and K. C. Yeh, Resonant and nonresonant wave-wave
interactions in an isothermal atmosphere, J. Gephys. Res.
\textbf{93}, 3729-3744 (1988).

\bibitem{Fritts1992}
D. C. Fritts, S. Sun, D.-Y. Wang, Wave-wave interactions in a
compressible atmosphere 1. A general formulation including
rotation and wind shear, J. Gephys. Res. \textbf{97}, 9975-9988
(1992).

\bibitem{Huang1991}
C.~S. Huang and J. Li, Weak nonlinear theory of the ionospheric
response to atmospheric gravity waves in the F-region, Journ.
Atmosphere and Terrest. Phys. \textbf{53}, 903-908 (1991).

\bibitem{Huang1992}
C.~S. Huang and J. Li, Interaction of atmospheric gravity solitary
waves with ion acoustic solitary waves in the ionospheric
F-region, Journ. Atmosphere and Terrest. Phys. \textbf{54},
951-956 (1992).

\bibitem{Fedun2013}
O. Onishchenko, O. Pokhotelov, and V. Fedun, Convective cells of
internal gravity waves in the earth's atmosphere with finite
temperature gradient, Ann. Geophys. \textbf{31}, 459-462 (2013).

\bibitem{Jovanovic2001}
D. Jovanovi\'{c}, L. Stenflo, and P. K. Shukla, Acoustic gravity
tripolar vortices, Phys. Lett. A \textbf{279}, 70-74 (2001).

\bibitem{Jovanovic2002}
D. Jovanovi\'{c}, L. Stenflo, and P. K. Shukla, Acoustic-gravity
nonlinear structures, Nonlin. Proc. Geophys. \textbf{9}, 333-339
(2002).

\bibitem{Shukla1998}
P.~K. Shukla and A. A. Shaikh, Dust-acoustic gravity vortices in a
nonuniform dusty atmosphere, Phys. Scripta \textbf{T75}, 247-248
(1998).

\bibitem{Fedun2016}
O. G. Onishchenko, W. Horton, O. A. Pokhotelov, and V. Fedun,
"Explosively growing" vortices of unstably stratified atmosphere,
J. Geophys. Res. Atmos. \textbf{121}, 11,264-11,268 (2016).

\bibitem{Fedun2021}
O. Onishchenko, V. Fedun, I. Ballai, A. Kryshtal and G. Verth,
Generation of localised vertical streams in unstable stratified
atmosphere, Fluids \textbf{6}, 454 (2021).

\bibitem{Misra2022IEEE}
A.~P. Misra, A. Roy, D. Chatterjee, T.~D. Kaladze, Internal
gravity waves in the Earth's ionosphere, IEEE Transactions in
Plasma Science \textbf{50}, 2603-2608 (2022).

\bibitem{Misra2022AdvSpace}
T.~D. Kaladze, A.~P. Misra, A. Roy, D. Chatterjee, Nonlinear
evolution of internal gravity waves in the Earth's ionosphere:
Analytical and numerical approach, Adv. Space Research
\textbf{69}, 3374-3385 (2022).

\bibitem{Lashkin2024}
V. M. Lashkin and O. K. Cheremnykh, Nonlinear internal gravity
waves in the atmosphere: Rogue waves, breathers and dark solitons,
Commun. Nonlinear Sci. Numer. Simulat. \textbf{130}, 107757
(2024).

\bibitem{Huang2014}
K. M. Huang, S. D. Zhang, F. Yi, C. M. Huang, Q. Gan, Y. Gong, and
Y. H. Zhang, Nonlinear interaction of gravity waves in a
nonisothermal and dissipative atmosphere, Ann. Gephys.
\textbf{32}, 263-275 (2014).

\bibitem{Fritts2015}
D. C. Fritts, B. Laughman, T. S. Lund, and J. B. Snively,
Self-acceleration and instability of gravity wave packets: 1.
Effects of temporal localization, J. Geophys. Res. Atmos.
\textbf{120}, 8783-8803 (2015).

\bibitem{Snively2017}
J. B. Snively, Nonlinear gravity wave forcing as a source of
acoustic waves in the mesosphere, thermosphere, and ionosphere,
Geophys. Res. Lett. \textbf{44}, 12020-12027 (2017).

\bibitem{Fritts2019}
T. Mixa, D. Fritts, T. Lund, B. Laughman, L. Wang, and L. Kantha,
Numerical simulations of high-frequency gravity wave propagation
through fine structures in the mesosphere, J. Geophys. Res. Atmos.
\textbf{124}, 9372-9390 (2019).

\bibitem{Turitsyn1985}
S. K. Turitsyn, Spatial dispersion of nonlinearity and stability
of multidimensional solitons, Theor. Math. Phys. \textbf{64}, 226
(1985).

\bibitem{Krolikovski2004}
W. Kr\'{o}likowski, O. Bang, N. I. Nikolov, D. Neshev, J. Wyller,
J. J. Rasmussen, and D. Edmundson, Modulational instability,
solitons and beam propagation in spatially nonlocal nonlinear
media, J. Opt. B: Quantum Semiclassical Opt. \textbf{6}, S288
(2004).

\bibitem{Torner2006}
C. Rotschild, M. Segev, Z. Xu, Y. V. Kartashov, L.~Torner, and
O.~Cohen, Two-dimensional multipole solitons in nonlocal nonlinear
media, Opt. Lett. \textbf{31}, 3312-3314 (2005).

\bibitem{Lashkin2006}
A. I. Yakimenko, V. M. Lashkin, and O. O. Prikhodko, Dynamics of
two-dimensional coherent structures in nonlocal nonlinear media,
Phys. Rev. E \textbf{73}, 066605 (2006).

\bibitem{Lashkin2007PLA}
V. M. Lashkin, A. I. Yakimenko, and O. O. Prikhodko,
Two-dimensional nonlocal multisolitons, Phys. Lett. A
\textbf{366}, 422-427 (2007).

\bibitem{Lashkin2007POP}
V. M. Lashkin, Two-dimensional ring-like vortex and multisoliton
nonlinear structures at the upper-hybrid resonance, Phys. Plasmas
\textbf{14}, 102311 (2007).

\bibitem{Lashkin2007PRA}
V. M. Lashkin, Two-dimensional nonlocal vortices, multipole
solitons, and rotating multisolitons in dipolar Bose-Einstein
condensates, Phys. Rev. A \textbf{75}, 043607 (2007).

\bibitem{Zakharov1972}
V.~E. Zakharov, Collapse of Langmuir waves, Sov. Phys. JETP
\textbf{35}, 908-914 (1972).

\bibitem{Zakharov_UFN2012}
V.~E. Zakharov and E.~A. Kuznetsov, Solitons and collapses: two
evolution scenarios of nonlinear wave systems, Physics-Uspekhi
\textbf{55}, 535-556 (2012).

\bibitem{Kivshar_book2003}
Y.~S. Kivshar and G.~P. Agrawal, \emph{Optical Solitons: From
Fibers to Photonic Crystals} (Academic Press, San Diego, 2003).

\bibitem{Dodd1982}
R.~K. Dodd, J.~C. Eilbeck, J. D. Gibbon, and H. C. Morris, \emph{
Solitons and Nonlinear Wave Equations} (Academic Press, London,
1982).

\bibitem{John1982}
F.~John, \emph{Partial Differential Equations}  (Springer-Verlag,
New York, 1982).

\bibitem{Rubenchik1986}
E.~A. Kuznetsov, A.~M. Rubenchik,  and V.~E. Zakharov, Soliton
stability in plasmas and hydrodynamics, Phys. Rep. \textbf{142},
103-165 (1986).

\bibitem{Lashkin2020}
V. M. Lashkin, Stable three-dimensional Langmuir vortex soliton,
Phys. Plasmas \textbf{27}, 042106 (2020).

\bibitem{Benney1977}
D. J. Benney, A general theory for interactions between short and
long waves, Stud. Appl. Math. \textbf{56}, 81-94 (1977).

\bibitem{Newell1978}
A. C. Newell, Long waves-short waves; a solvable model, SIAM J.
Appl. Math. \textbf{35}, 650-664 (1978).

\bibitem{Kanna2014}
T. Kanna, M. Vijayajayanthi, and M. Lakshmanan, Mixed solitons in
a (2+1)-dimensional multicomponent long-wave-short-wave system,
Phys. Rev. E \textbf{90}, 042901 (2014).

\bibitem{Hobart1963}
R. H. Hobart, On the instability of a class of unitary field
models, Proc. Phys. Soc. (London) \textbf{82}, 201-203 (1963).

\bibitem{Derrick1964}
G. H. Derrick, Comments on nonlinear wave equations as models for
elementary particles, J. Math. Phys. \textbf{5}, 1252-1254 (1964).

\bibitem{Makhankov1993}
 V. G. Makhankov, Yu. P. Rybakov, V. I. Sanyuk,
\emph{The Skyrme Model: Fundamentals, Methods, Applications}
(Springer-Verlag, New York, 1993).

\bibitem{Sulem1999}
C.~Sulem, P.-L. Sulem, \emph{The nonlinear Schr\"{o}dinger
equation: self-focusing and wave collapse} (Springer-Verlag, New
York, 1999).

\end{thebibliography}
\end{document}